\documentclass[conference]{IEEEtran}
\IEEEoverridecommandlockouts
\usepackage{cite}
\usepackage{amsmath,amssymb,amsfonts}
\usepackage{algorithmic}
\usepackage{graphicx}
\usepackage{textcomp}
\usepackage{xcolor}
\def\BibTeX{{\rm B\kern-.05em{\sc i\kern-.025em b}\kern-.08em
		T\kern-.1667em\lower.7ex\hbox{E}\kern-.125emX}}
\begin{document}
	
	\title{A Natural Lane Changing Decision Model For Mixed Traffic Flow Based On Extreme Value Theory\\
		\thanks{*This work was supported by Supported by the Fundamental Research Funds for the Central Universities(2023YJS130), The National Natural Science Foundation of China (52272328) and Beijing Municipal Natural Science Foundation (L211022).}
	}
	
	\author{\IEEEauthorblockN{1\textsuperscript{st} Jiali Peng}
		\IEEEauthorblockA{\textit{School of Electronic and Information Engineering} \\
			\textit{Beijing Jiao tong University}\\
			Beijing, China \\
			jialipeng@bjtu.edu.cn}
		\and
		\IEEEauthorblockN{2\textsuperscript{nd} Wei Shangguan}
		\IEEEauthorblockA{\textit{School of Electronic and Information Engineering} \\
			\textit{Beijing Jiao tong University}\\
			Beijing, China \\
			wshg@bjtu.edu.cn}
		\and
		\IEEEauthorblockN{3\textsuperscript{th} Rui Luo}
		\IEEEauthorblockA{\textit{School of Electronic and Information Engineering} \\
			\textit{Beijing Jiao tong University}\\
			Beijing, China \\
			21111066@bjtu.edu.cn}
		\and
		\IEEEauthorblockN{4\textsuperscript{th} Ke Gao }
		\IEEEauthorblockA{\textit{ School of Electronic and Information Engineering} \\
			\textit{Southwest Jiaotong University}\\
			Chengdu, China \\
			1194170132@qq.com}
	}
	
	\maketitle
	
	\begin{abstract}
With the high frequency of highway accidents, studying how to use connected automated vehicle (CAV) to improve traffic efficiency and safety will become an important issue. In order to investigate how CAV can use the connected information for decision making, this study proposed a natural lane changing decision model for CAV to adapt the mixed traffic flow based on extreme value theory. Firstly, on the bias of the mixed vehicle behavior analysis, the acceleration, deceleration, and randomization rules of the cellular automata model of mixed traffic flow in two lanes are developed. Secondly,the maximum value of CAV's lane change probability at each distance by extreme value distribution are modeled. Finally, a numerical simulation is conducted to analyze the  trajectory-velocity  diagram of mixed traffic flow, average travel time and average speed  under different penetration rates of CAV. The result shows that our model can avoid the traffic risk well and significantly improve traffic efficiency and safety.
	\end{abstract}
	

\section{INTRODUCTION}

\subsection{Overview}
The last decade has witnessed the evolution and advanced of new technologies for Intelligent Transportation Systems (ITS)\cite{c1}. The most technologies  are deployed as part of Information Technologies systems aiming to improve road safety, driver comfort, transport efficiency and conduct to refinements in secondary effects at environmental level as well as in energy management\cite{c2}. Among them The emergence of connected and automated vehicle (CAV)  has given a unique opportunity to improve mobility through vehicle-to-vehicle (V2V) and vehicle to-infrastructure (V2I) communications\cite{c3}. CAV can collect environmental information about local traffic, weather and infrastructure using different technologies such as cameras, lasers, sensors, radar, etc\cite{c4}. The collected information is then shared between the individual groups of vehicles through V2V or V2I communications to facilitate efficient driving. On the other hand, the same collected data may be processed by roadside units(RSU) and communicated to different CAV either directly or via a public network to improve traffic management\cite{c5}. However, relevant reports show that the penetration rate of CAV on the road can only reach $24.8\%$ in 2045\cite{c6}, indicating that with the current high utilization of human-driven vehicles (HDV), CAV and HDV will coexist on the road in the long term, and the future traffic flow will be randomly mixed, which is also called mixed traffic environment\cite{c7}. This gives rise to a significant challenge for avoiding accidents of CAV in a limited length of highway. When risky situations such as accidents, weather and road conditions or specific events occur. Messages of this nature  about them will be broadcast towards CAV currently using a road network about emerging situations that could have potential impact on the traffic condition. The information is propagated upstream from the location of a specific situation (accident, work zones, slippery road, adverse weather conditions, etc.) with the support of technology infrastructure\cite{c8}.

Our paper proposes a natural lane changing decision model for CAV to adapt the mixed traffic flow based on extreme value theory. Specifically, the motion decisions of the  CAV act as constraints and guidance to the other CAV and HDV. It will increase the average speed as to avoid traffic jam, and decrease the average time and distance in the risk zone. And there is a growing interest in the application of extreme value theory(EVT) in the recent literature. It provides a single dimension to identify crashes that can well fit in the hierarchy of safety pyramid. Due to this unique feature of EVT, this study adopts EVT to establishing mandatory lane-changing and speed guidance process for CAV, which is called 'natural'.

\begin{figure*}
	\centering
	\includegraphics[scale=0.6]{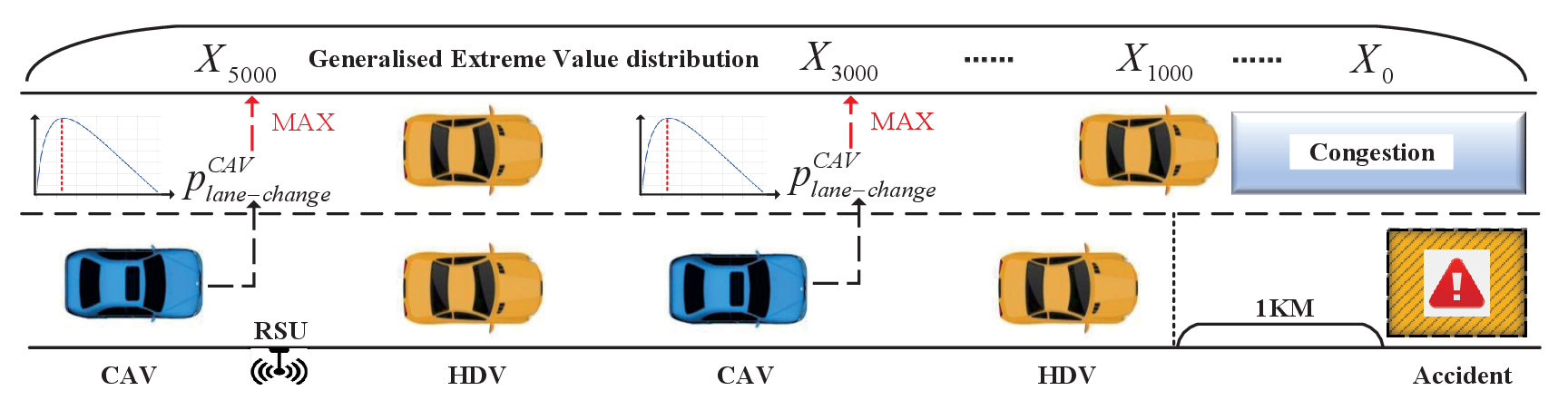}
	\caption{The natural lane changing decision model of CAV under mixed traffic flow}
	\label{figurelabel}
\end{figure*}

\subsection{Related Works}
A review of literature indicate that mixed traffic flow modeling can be classified into numerical simulation-based studies and cellular automata-based studies. This section covers some representative studies related to cellular automata-based studies in the ensuing paragraphs.

As an important tool of microscopic traffic flow simulation, cellular automata (CA) can simulate complex traffic phenomena based on some simple rules. Nagel and Schrekenberg\cite{c9} proposed the classic NS model. With the development of CAV technology and the high efficiency of CA model in computer operations. In addition, as alluded above, the popularization of CAV will take a long time, so the impact of it can only be analyzed through simulation. Some scholars have gradually studied the mixed traffic flow based on CA in the intelligent transportation system. These studies involve traffic flow characteristics (including single-lane and multi-lane), intersection traffic control, traffic safety issues, and public transportation.

On the issue of traffic efficiency. For example, Zhao et al respectively proposed  the CE-NS model and the CI-NS model. The CI-NS model considered the communication between CAV. The results showed that CAV can significantly optimize the traffic flow at intersections only when the number of CAV exceeds a certain percentage\cite{c10};Jiang et al proposes a cellular automata model of mixed traffic flow considering the driving behavior of the platoon of CAV. The penetration rate of CAV and the maximum platoon size increase can improve the road capacity\cite{c11}.

On the issue of traffic safety. For example, Chain and Wong compared CA and surrogate safety assessment models (SSAM) in terms of safety assessment. They showed that CA model can replicate realistic conflicts at signalized intersection\cite{c12}. Marzoug et al propose a two-lane cellular automata model that explains the relationship between traffic-related parameters and likelihood of accidents at a signalized intersection. The model results illustrated that the traffic at the intersection is more dangerous adopting asymmetric lane-changing rules than symmetric ones\cite{c13}.

The contribution of this study is twofold. In Fig.1.First, by modeling and simulation of traffic accidents  at highway using cellular automata, this study explains how Impact of traffic accidents on the upstream of the road varies in a mixed traffic environment as a function of time and different CAV penetration rate. Second, as one of the first studies on the application of Extreme Value Theory for  decision-making of CAV in a mixed traffic environment, this study provides valuable insights into traffic risk when accidents for a connected environment are occured which can better guide mixed traffic flow, thus reducing the severity of traffic congestion caused by traffic accidents. Such findings will help us to better understand the impact of a mixed traffic environment on diverse vehicles and suggest natural decision for a group of CAV who facing traffic risks.

\section{METHODOLOGY}

\subsection{Cellular automaton model of mixed traffic flow}
The CA model can simulate complex traffic phenomena based on some simple rules. For the characteristics of the mixed traffic which mainly contains  HDV and CAV, corresponding rules are developed respectively. Finally, the CA model of mixed traffic flow is obtained.
To distinguish different car-following and lane-changing modes, let ${{\alpha }_{n}}$ and ${{\beta }_{n}}$ are $0-1$ variable used to judge whether the  type of the vehicle n is HDV and CAV, respectively. 
$$
{{\alpha }_{n}}=\left\{ \begin{array}{*{35}{l}}
	1,\textnormal{ if the type of the vehicle }n\textnormal{ is HDV }  \\
	0,\textnormal{ else },  \\
\end{array} \right.\eqno{(1)}
$$
$$
{{\beta }_{n}}=\left\{ \begin{array}{*{35}{l}}
	1,\textnormal{ if the type of the vehicle }n\textnormal{ is CAV }  \\
	0,\textnormal{ else }  \\
\end{array} \right.\eqno{(2)}
$$
$$
\alpha_n+\beta_n=1\eqno{(3)}
$$

When the driving behavior of the leading vehicle changes, the HDV needs time to perceive, recognize and judge the change in the driving state of the leading vehicle before taking measures; this time is called reaction time $\tau$ in this study. Besides, the vehicle may produce random deceleration because of uncertain factors such as the driver's mentality. The CAV uses the on-board sensing system or V2V/V2I technologies to obtain the status information of the leader. Therefore, it can quickly capture the change of the leader's behavior and take corresponding measures. The reaction time is the processing time of the on-board sensing system, which is shorter than that of the HDV.

Hence, before proceeding with the design of the CA model rules, it is necessary to introduce the concept of safety distance, which is extended from the Gipps model. It is the minimum distance that the follower does not collide with the leader vehicle when the leader vehicle brakes suddenly. The safety distance of the vehicle defined in this way can ensure that the vehicle is safe in all cases. According to Newton's second law, we know that the safe distance between adjacent vehicles in mixed traffic flow can be determined by
$$
d_{n}^{safe}={{v}_{n}}(t)\left( {{\alpha }_{n}}{{\tau }^{\textnormal{HDV}}}+{{\beta }_{n}}{{\tau }^{\textnormal{CAV}}} \right)
$$
$$
+\left( {{\alpha }_{n}}+{{\beta }_{n}} \right)\left( \frac{{{v}_{n}}{{(t)}^{2}}-{{v}_{n-1}}{{(t)}^{2}}}{2B} \right)\eqno{(4)}
$$

where $d_{n}^{safe}$ is the safe distance between the vehicle $n$ and vehicle $n-1$. $v_n(t)$ is the speed of the vehicle $n$ at time $t$ (m/s). $\tau^{\mathrm{HDV}}$ and $\tau^{\mathrm{CAV}}$ are the reaction time (s) of the HDV and the CAV, respectively. $B$ is the maximum deceleration of the vehicle. Then based on the safe distance of vehicles with different type,  the acceleration, deceleration, randomization, and position update rules of the CA model of mixed traffic flow are proposed as follows.

(a) Acceleration

For HDV and CAV, when the distance$d_n$between vehicle$n$and its leading vehicle$n-1$is greater than the safety distance $d_{n}^{safe}$, vehicle$n$will accelerate due to the pursuit of higher speed. To meet the rationality of simulation acceleration, the speed of vehicle$n$at the next time step is the minimum of $v_n(t)+\alpha_n \Delta t, V_{\max }$, and $d_n / \Delta t$.
$v_n(t+\Delta t)=\left(\alpha_n+\beta_n\right) \min \left(v_n(t)+a \Delta t, V_{\max }, d_n / \Delta t\right)$
$d_n=x_{n-1}(t)-x_n(t)-l_{n-1}$,
where $\Delta t$ is the time step.${{v}_{n}}(t+\Delta t)$ is the speed of vehicle $\mathrm{n}$ at time $t+\Delta t.a$ is the acceleration of the vehicle ${{V}_{\max }}$ is the maximum speed of the vehicle ${{d}_{n}}$ is the distance between vehicle $n$ and vehicle $n-1$. $v_{n-1}(t+\Delta t)$ is the speed of vehicle $n-1$ at time $t+\Delta t$.
$d_{n}^{safe}$is the safe distance of vehicle $n$, which can be obtained by Eq. (4). $x_{n-1}(t)$ is the position of vehicle $n-1$ at time $t.{{x}_{n}}(t)$ is the position of vehicle $n$ at time $t$. $l_{n-1}$ is the length of vehicle $n-1$.

(b) Deceleration

When the distance dn between the vehicle $n$ and the vehicle $n-1$ is less than or equal to the safe distance $d_{n}^{safe}$. the vehicle will decelerate to ensure driving safety. Therefore, the speed of vehicle $n$ at the next moment is $v_{n-1}(t+\Delta t)$. The form of the deceleration rule is 
$$
{{v}_{n}}(t+\Delta t)=\left( {{\alpha }_{n}}+{{\beta }_{n}} \right)\min \left( {{v}_{n}}(t),{{d}_{n}}/\Delta t \right)\eqno{(5)}
$$

(c) Randomization

For better simulation of HDV, random slowing probability $p_{slow}^{{}}$is introduced in this study, and the vehicle will be slowed down according to the random probability. The random deceleration state of the vehicle is the same in the same reaction time and all HDV in each simulation step will make this determination. Since there is no instability in the CAV, it is specified that there is no randomization process. The randomization rule for HDV is obtained by the equation
$$
{{v}_{n}}(t+\Delta t)=\max \left( {{v}_{n}}(t)-b\Delta t,0 \right)\eqno{(6)}
$$
where $b$ is the random deceleration of the vehicle.$p_{slow }$is the randomization probability.

(d) Lane-changing

Drivers are always changing lanes to maintain the maximum speed possible or to avoid some accidents. Usually, lane change rules can be symmetrical or asymmetrical rules related to vehicles or lanes. The symmetric rules consider both lanes and all vehicles equally.  and in addition, lane change is allowed only when certain safety conditions are satisfied. 
This paper uses symmetric rules which they can change the lane if they satisfied the lane-changing rules. All steps are applied to all vehicles including CAV and HDV in parallel manner. Here, vehicles can change the lane according to the following symmetric rules (with respect to the vehicles kinds as well as with respect to the lanes):
$$
\left\{ \begin{array}{*{35}{l}}
		 rand\le p_{lane-change}^{HDV/CAV}\\
		\\
		 gap_{\inf ront}^{current}\left( t \right)<{{v}_{n}}\left( t \right)+a\Delta t\\
		\\
		 gap_{\inf ront}^{other}\left( t \right)>gap_{\inf ront}^{current}\left( t \right)\\
		\\
		 v_{n-1}^{other}\left( t \right)<g_{back}^{other}\left( t \right)\textnormal{ }\\

\end{array} \right.		\eqno{(7)}
$$
Where:
$gap_{\inf ront}^{current}\left( t \right)$ and $gap_{\inf ront}^{other}\left( t \right)$ are the gap in front in the current lane (left or right) and the other lane at the step t, respectively.
$g_{back}^{other}\left( t \right)$ is the back gap in the other lane (left or right) at the step t.
$v_{n-1}^{other}\left( t \right)$ is the maximum velocity of the vehicle behind in the other lane at the step t.
$rand\le p_{lane-change}^{HDV/CAV}$ is the lane-changing probability of HDV and CAV.
If the above lane-changing rules are satisfied, then the vehicle can change the lane.

(e)Position update

After the next time step's speed and lane is updated, the position of the vehicle is updated
$$
x_n(t+\Delta t)=x_n(t)+v_n(t+\Delta t) \cdot \Delta t\eqno{(8)}
$$

\subsection{EVT models for CAV lane changing decision}

With the pioneering work of Tarko et al\cite{c14}, EVT paved its way in traffic safety analysis about two decades ago. With some preliminary analysis on estimating crash risk using traffic conflicts, the efficacy of EVT was confirmed in that study. Several attempts have been made in recent years to obtain real trajectory data in a connected environment and assess safety\cite{c15}\cite{c16}. 

The synthesis of the literature suggests that majority of the studies employed EVT to estimate crash risk probability. The application of EVT seems to be most relevant in this case because of its capability of estimating crash risk without using historical crash records.

However, the real-time application factor is missing in the existing EVT model, which serves as the primary tool for studying the probability of extreme events and is able to adequately represent the extreme variability of random variables. Only through real-time application can we gain insight into the risky behavior of traffic subjects and reduce such risks, thus improving the behavioral reliability and safety of these vehicles, futhermore, and helping to improve our understanding of the relationship between traffic environment and traffic safety. These issues motivate the study in this paper.

\begin{figure}[h]
	\centering
	\includegraphics[scale=0.65]{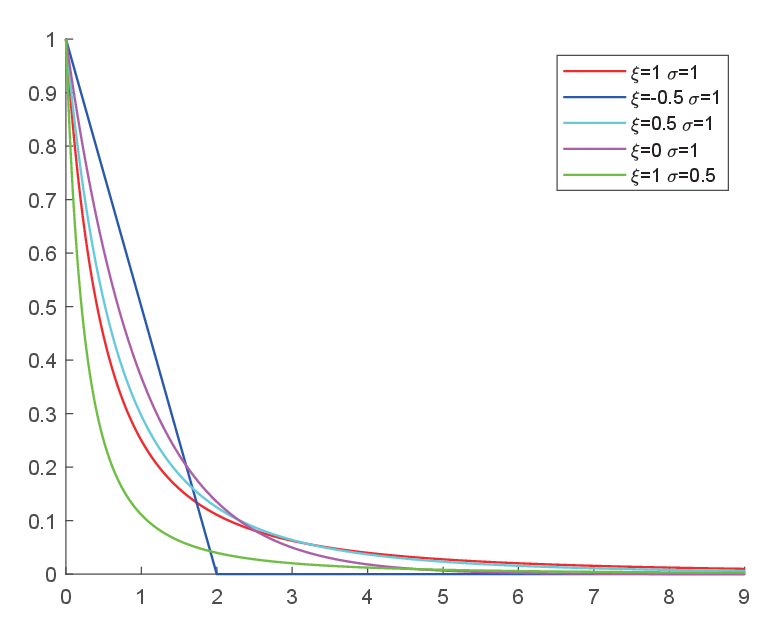}
	\caption{Extreme value distribution under different parameter values}
	\label{figurelabel}
\end{figure}

When a CAV is driving on a highway where it has been informed of an accident ahead, the traffic risk will increase with proximity to the accident site, but may not be linearly approximated. Therefore, in order to adapt natural driving behavior, the risk-distance relationship can be estimated to determine when the CAV should make a lane change to induce the HDV, to ensure maximum safety and efficiency, and may to avoid traffic congestion or even secondary accidents due to accidents.To this end, Generalised Pareto distribution is applied because of its upper tail that deals with extreme values. More specifically, consider the maximum $M_n=\max \left(X_1, X_2, \cdots, X_n\right)$ . Among them, we define $X$ as the maximum value of CAV lane change probability $p_{lane-change}^{CAV}$ at different distances from the accident site. For example, at 1km from the accident site, the possibility that CAV can bear the maximum risk (lane change avoidance) in the face of accidents ahead is ${{X}_{1000}}$. And this definition is made because the closer the distance to the accident site, the greater the corresponding traffic risk (vehicle driving safety, traffic congestion, etc.), which can also be characterized as an increasingly extreme and less probable event, because as a CAV it has the ability to obtain a large amount of information, for example, it is basically impossible for a CAV to slow down urgently at 100m before the accident and then change lanes to avoid the accident, the probability is very small. Then if a sequence of constants $a_n>0$ and $b_n>0$ exists, then ${Pr}\left[\frac{\left(M_n-b_n\right)}{a_n \leq z}\right] \rightarrow G(z)$ as $n \rightarrow \infty$ for a nondegenerate distribution function $G$ , then $G$ belongs to the Generalised Extreme Value family, where the distributional function is
$$
G(z)=\exp \left\{-\left[1+\xi\left(\frac{z-\mu}{\sigma}\right)\right]^{-1 / \xi}\right\}\eqno{(9)}
$$
defined on $z: 1+\xi\left[\frac{z-\mu}{\sigma}\right]>0$, as shown in Fig. 2, where, $-\infty<\mu<\infty$ indicates the location parameter, $\sigma>0$ denotes the scale parameter, and $\xi$ represents the shape of a Generalised Extreme Value distribution. As a result of assuming that the parent distribution ${Pr}\left[\frac{\left(M_n-b_n\right)}{a_n \leq z}\right] \rightarrow G(z)$ is known, the distribution of lane change probablity would also be know,then we can use the $rand\le p_{lane-change}^{CAV}$ to make CAV perform lane change.

\begin{figure*}[h]
	\centering
	\begin{tabular}{ccc}
		\includegraphics[width=6cm]{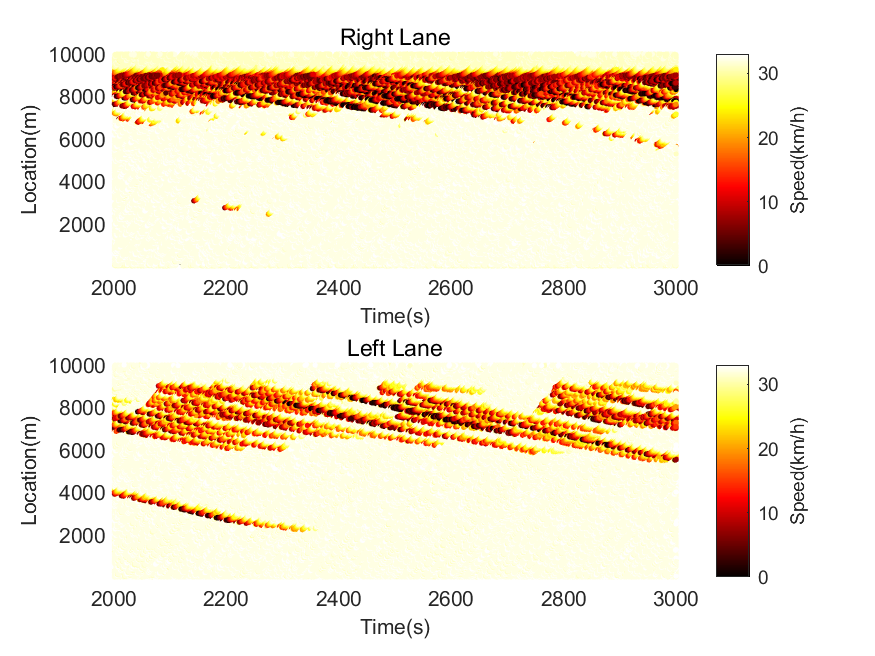} &    \includegraphics[width=6cm]{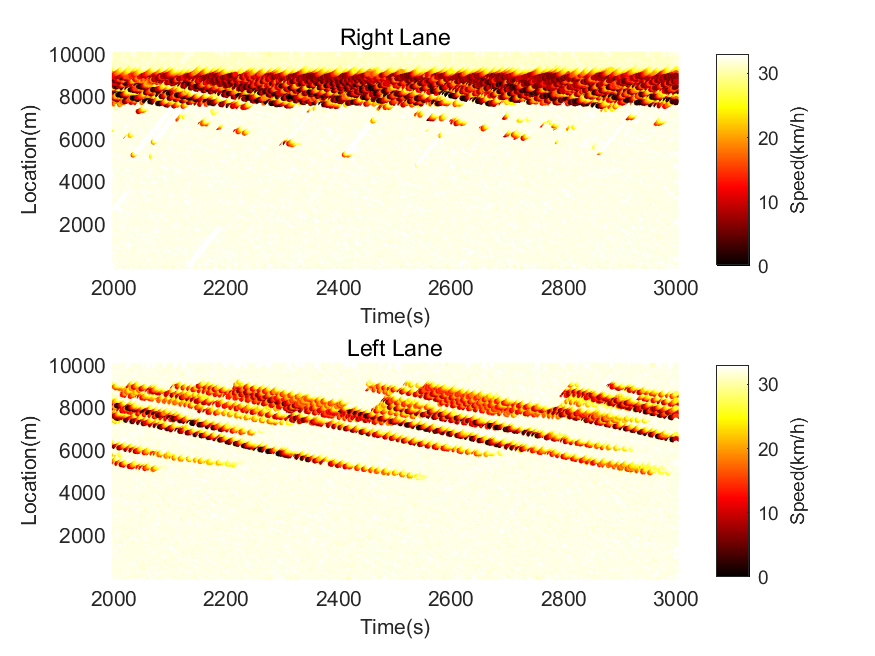}  &    \includegraphics[width=6cm]{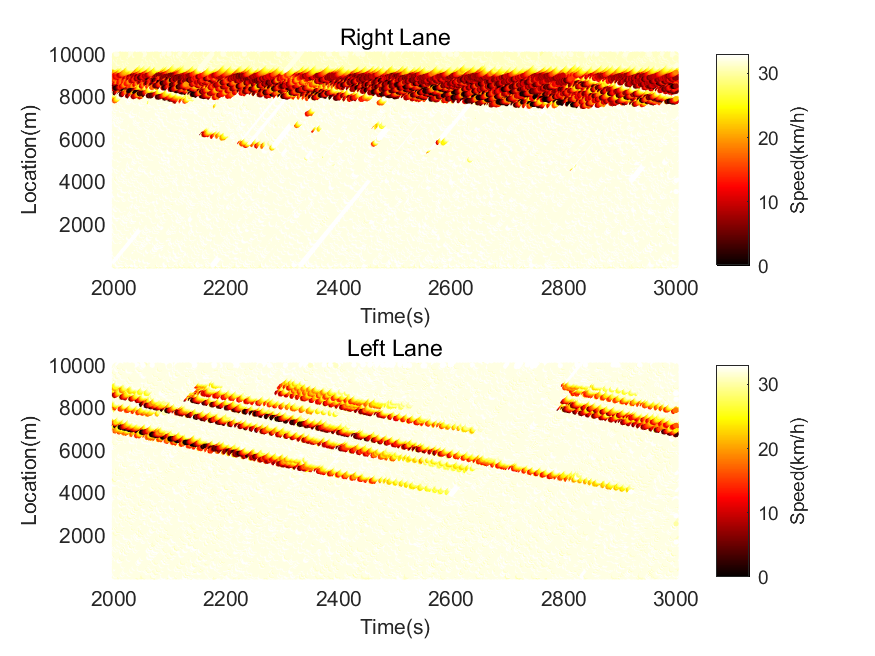}      \\
		(a) 0\% & (b) 20\% & (c) 40\%\\
		\includegraphics[width=6cm]{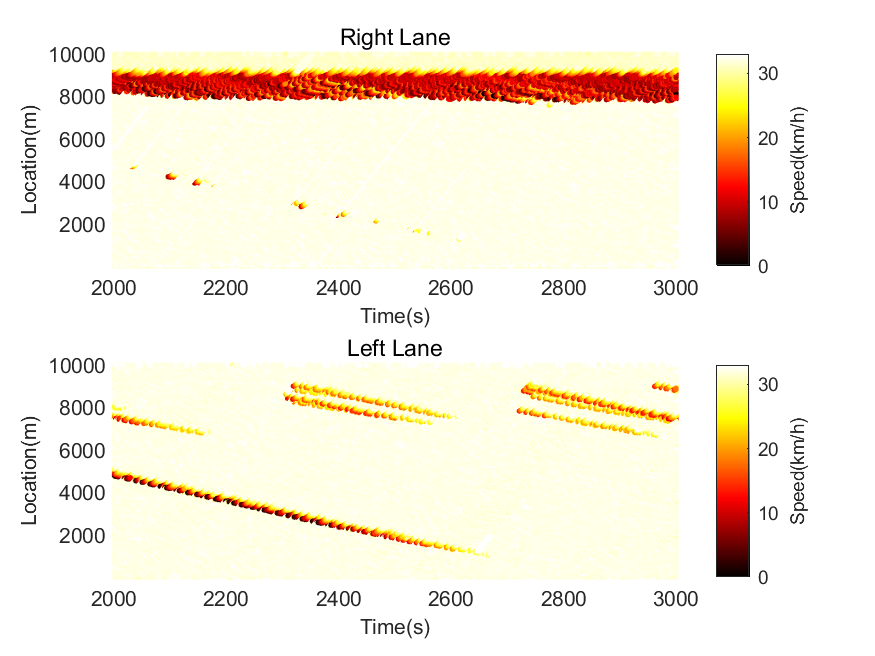} &    \includegraphics[width=6cm]{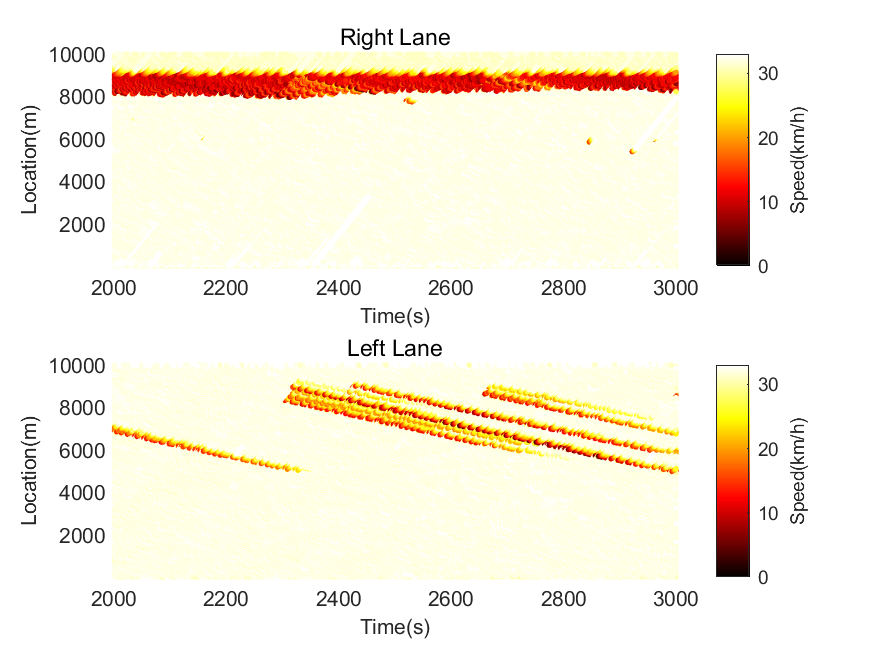}   &    \includegraphics[width=6cm]{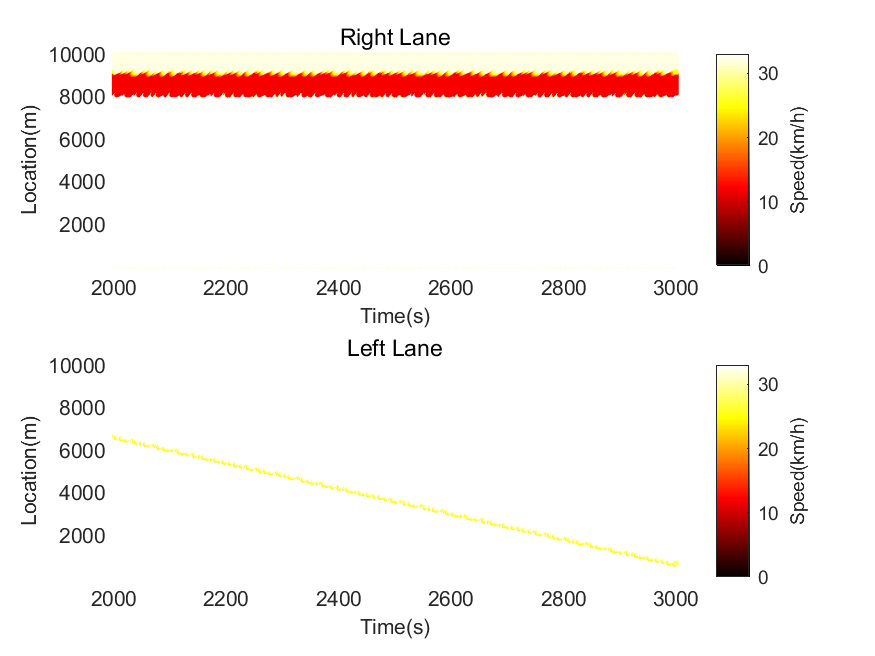}     \\
		(d)60\% & (e) 80\% & (f) 100\%\\
	\end{tabular}
	\caption{Trajectory-velocity diagram of mixed traffic flow}
	\label{figurelabel}
\end{figure*}

\begin{table}[h]
	\centering
	\caption{Parameters for the simulation}
	\renewcommand{\arraystretch}{1.4}
	\scalebox{1}{
		{\begin{tabular}{cc} 
				
				\hline
				Parameter&Value\\
				\hline
				$\xi$&1\\
				$\mu$&0\\
				$\sigma$&1\\
                $Vmax$&$330 cells/step$($33 m/s$)\\
                $a$&$30 cells/step^2$ ($3 m/s^2$)\\
                $b$&$30 cells/step^2$ ($3 m/s^2$)\\
                Road Length&$10^5 cells(10km)$\\
                Accident Site&$9\times10^4 cells(9km)$\\
                Vehicle Length&50 $cells(5m)$\\
                Step Length&3000 steps\\
                Timestep&0.1s\\
                $p_{slow}$&0.2\\
                $p_{lane-change}^{HDV}$&0.07\\
                ${{\tau }^{\textnormal{HDV}}}$&20 steps(2s)\\
                ${{\tau }^{\textnormal{CAV}}}$&6 steps(0.6s)\\   
                Penetration rates&0-100\%\\             
				\hline
	\end{tabular}}}
\end{table}

\section{SIMULATION\&RESULTS}

In this paper, a mixed traffic environment on a two-lane highway is constructed using numerical simulation, and the usability of the proposed CAV natural lane changing decision model is verified based on this simulation environment. In the numerical simulation, the lane consists of $10^5$ cells, each cell length is 0.1 m, so the road length is L=10km. The traffic accident occurs in the right lane at 10km, where it is impassable, and the vehicle needs to change lanes in advance to avoid the accident when it is observed, where the observation distance of HDV is 1km according to the forward-looking effect, that is, HDV can observe the traffic accident in front of it at 9km. The HDV can observe the traffic accident in front of it at 9km, and then prepare to start the next operation. The simulation time step is set to 0.1 s, and the total step length is 3000. the last 1000 steps are recorded to investigate the traffic flow characteristics. The simulation uses the periodic boundary. In the initial state, the vehicles are uniformly distributed on the road and the speed is generated randomly. The vehicle length is 50 cells (5 m). The speed limit is Vmax=330 cells/step (the general speed limit of Chinese highways is 120 km/h). The general acceleration, random deceleration and maximum deceleration of the vehicles are a=$30 cells/step^2$ ($3 m/s^2$), b=30 cells/step1 ($3 m/s^2$) and b=$50 cells/step^2$ ($5 m/s^2$), respectively, and the randomization probability of HDV is set to 0.2. In the simulation, the response times of HDV and CAV are set to 20 steps (2 s) and 6 steps (0.6 s).The values of parameters in the numerical simulation are shown in Table 1. In order to study the influence of the penetration rate of CAVs on the characteristics of mixed traffic flow, the penetration rates are set to  0\%, 20\%, 40\%, 60\%, 80\% and 100\%.

\subsection{Trajectory-velocity Diagram Analysis} 

Figure 3 shows the speed and trajectory of vehicles on two lanes with different CAV penetration rates at a traffic density of 50 vehicles/km, where the darker color indicates lower speed, which means greater traffic congestion. From Fig. 3(a)-(f), it can be seen that as the penetration rate increases, the congestion area on the two lanes becomes smaller, the distance of congestion propagation becomes shorter, the impact area is smaller, and the congestion in the right lane where the accident occurs will be relieved. As shown in Figure 3(a), when the penetration rate of CAV is 0\%, the congestion due to the traffic accident in the right lane propagates to almost the whole road section, causing a substantial traffic efficiency reduction. As shown in Figure 3(f), when the penetration of CAV is 100\%, traffic congestion exists only in the accident area of the right lane, and there is no traffic congestion in other areas. The main reason is that the CAV decision of applying extreme value distribution proposed in this paper still has a very small number of CAVs to make speed reduction and lane change only before 10 KM (where the accident occurred), thus causing traffic congestion, and the rest of CAV can all adjust quickly during the driving process and then drive at the highest speed.

\begin{table}[t]
	\centering
	\caption{Average travel time and speed of mixed traffic flow}
	\renewcommand{\arraystretch}{1.2}
\scalebox{0.8}{\begin{tabular}{|c|ccc|} 
				
				\hline
				\textbf{Penetration Rates} & \textbf{\textit{Vehicle Type}}& \textbf{\textit{Average Travel Time(s)}}& \textbf{\textit{Average Speed(m/s)}} \\
				\hline
				
				&CAV&0&0\\
				
				0\%&HDV&395.223&25.486\\
				&ALL&395.223&25.486\\
				
				\hline
				
				&CAV&365.580&27.196\\
				
				20\%&HDV&376.451&26.399\\
				&ALL&374.277&26.559\\
				
				\hline
				
				&CAV&355.080&27.992\\
				
				40\%&HDV&367.267&27.127\\
				&ALL&362.392&27.473\\
				
				\hline
				
				&CAV&346.958&28.787\\
				
				60\%&HDV&348.738&28.719\\
				&ALL&347.670&28.760\\
				
				\hline
				
				&CAV&338.439&29.485\\
				
				80\%&HDV&349.145&28.578\\
				&ALL&340.580&29.303\\
				\hline
				&CAV&330.346&30.148\\
				
				100\%&HDV&0&0\\
				&ALL&330.346&30.148\\
				\hline
\end{tabular}}
\end{table}

\subsection{Traffic Efficiency Analysis}
In this paper, the average travel time and average speed of vehicles with different CAV penetration rates are calculated for a traffic density of 50 vehicles/km, and the results are shown in Table 2. Table 2 shows that as the penetration rate keeps increasing, the average travel time of all vehicles keeps getting shorter and the average speed keeps increasing. Among them, CAV and HDV also lead to shorter average travel time and higher average speed as the CAV penetration rate increases, mainly because all vehicles form a more stable flow state under the influence of the CAV decision method in this paper, which continuously reduces the impact of congestion propagation caused by traffic accidents. As can be seen from the table, when the penetration rate is 100\%, the average travel time is reduced by 16.42\% and the average speed is increased by 18.29\% compared with the pure HDV. When the penetration rate is 60\%, the average travel time of CAV is reduced by 12.21\% and the average speed is increased by 12.95\% compared to pure HDV. the average travel time of HDV is reduced by 11.76\% and the average speed is increased by 12.69\% compared to pure HDV. Combined with the above analysis, the following conclusions can be drawn: the large-scale application of CAV can effectively reduce traffic congestion, increase the efficiency of traffic and improve the driving experience of HDV.

\section{CONCLUSIONS}

This study proposed a natural lane changing decision model for CAV to adapt the mixed traffic flow based on extreme value theory. We designed to model the maximum value of CAV's lane change probability at each distance by extreme value distribution, and evaluated our method by setting different CAV penetration rates in a mixed traffic environment. The results showed that the model can better improve the average travel time and average vehicle speed, and CAV can also improve the travel efficiency of HDV. 
This work only considered the CAV lane change rule setting through a fixed extreme value distribution. In the future, a more realistic extreme value distribution will be obtained by fitting the natural driving dataset.  Futhermore, It is not enough to model the lane change probability as an extreme value distribution only, we will continue to explore the application of extreme value theory in the CAV decision making process to improve the operational safety and efficiency of the mixed traffic environment.


\begin{thebibliography}{99}

\bibitem{c1} Lian Y, Zhang G, Lee J, et al. Review on big data applications in safety research of intelligent transportation systems and connected/automated vehicles. Accident Analysis\&Prevention, vol. 146, pp. 105711, October 2020.
\bibitem{c2} Sumalee A, Ho H W. Smarter and more connected: Future intelligent transportation system. Iatss Research, vol. 42(2), pp. 67-71, July 2018.
\bibitem{c3} Ndashimye E, Ray S K, Sarkar N I, et al. Vehicle-to-infrastructure communication over multi-tier heterogeneous networks: A survey. Computer networks, vol. 112, pp. 144-166, January 2017.
\bibitem{c4} Guerrer J, Zeadally S, Contreras-Castillo J. Sensor technologies for intelligent transportation systems. Sensors, vol. 18(4), pp. 1212, April 2018.
\bibitem{c5} Xiong B K, Jiang R, Li X. Managing merging from a CAV lane to a human-driven vehicle lane considering the uncertainty of human driving. Transportation research part C: emerging technologies, vol. 142, pp. 103775, September 2022.
\bibitem{c6} Van Arem B, Van Driel C J G, Visser R. The impact of cooperative adaptive cruise control on traffic-flow characteristics. IEEE Transactions on intelligent transportation systems, vol. 7, pp. 429-436, December 2006.
\bibitem{c7} Peng J, Shangguan W, Chai L. Strategy of lane-changing coupling process for connected and automated vehicles in mixed traffic environment. Transportmetrica B: Transport Dynamics, December 2022, to be published.
\bibitem{c8} Ladino A, Laharotte P A, El Faouzi N E. System Level Impacts of V2I-based Speed Control Strategies: the SCOOP@ F project deployment scenarios. International Symposium on Transportation and Data Modeling. Ann Arbor, 2021.
\bibitem{c9} Nagel K, Schreckenberg M. A cellular automaton model for freeway traffic. Journal de physique I, vol. 2, pp. 2221-2229, December 1992.
\bibitem{c10} Zhao H T, Liu X R, Chen X X, et al. Cellular automata model for traffic flow at intersections in internet of vehicles. Physica A: Statistical Mechanics and its Applications, vol. 494, pp. 40-51, March 2018.
\bibitem{c11} Jiang Y, Wang S, Yao Z, et al. A cellular automata model for mixed traffic flow considering the driving behavior of connected automated vehicle platoons. Physica A: Statistical Mechanics and its Applications, vol. 582, pp. 126262, November 2021.
\bibitem{c12} Chai C, Wong Y D. Comparison of two simulation approaches to safety assessment: cellular automata and SSAM. Journal of Transportation Engineering, vol. 141, pp. 05015002, June 2015.
\bibitem{c13} Marzoug R, Lakouari N, Ez-Zahraouy H, et al. Modeling and simulation of car accidents at a signalized intersection using cellular automata. Physica A: Statistical Mechanics and its Applications, vol. 589, pp. 126599, March 2022.
\bibitem{c14} Songchitruksa P, Tarko A P. The extreme value theory approach to safety estimation. Accident Analysis \& Prevention, vol. 38, pp. 811-822,  July 2006.
\bibitem{c15} Ali Y, Haque M M, Zheng Z. An Extreme Value Theory approach to estimate crash risk during mandatory lane-changing in a connected environment. Analytic methods in accident research, vol. 33, pp. 100193, March 2022.
\bibitem{c16} Arun A, Haque M M, Bhaskar A, et al. A bivariate extreme value model for estimating crash frequency by severity using traffic conflicts. Analytic methods in accident research, vol. 32, pp. 100180, December 2021.
\bibitem{c17} Peng J, Shangguan W, Zhang L, et al. An Optimal scheduling method using multi-agent A* for Autonomous shuttle bus.2021 40th Chinese Control Conference (CCC). IEEE, Shanghai, 2021: 6040-6045.
\bibitem{c18} Shangguan W, Luo R, Song H, et al. High-Speed Train Platoon Dynamic Interval Optimization Based on Resilience Adjustment Strategy. IEEE Transactions on Intelligent Transportation Systems, vol. 23, no. 5, pp. 4402-4414, May 2022.
\bibitem{c19} Arun A, Haque M M, Bhaskar A, et al. A systematic mapping review of surrogate safety assessment using traffic conflict techniques. Accident Analysis \& Prevention, vol. 153, pp. 106016, April 2021.
\bibitem{c20} Cavadas J, Azevedo C L, Farah H, et al. Road safety of passing maneuvers: a bivariate extreme value theory approach under non-stationary conditions. Accident Analysis \& Prevention, vol. 134, pp. 105315, January 2020.
\bibitem{c21} Arun A, Haque M M, Washington S, et al. A physics-informed road user safety field theory for traffic safety assessments applying artificial intelligence-based video analytics. Analytic Methods in Accident Research, vol. 37, pp. 100252, March 2023.
\bibitem{c22} Wang C, Xu C, Xia J, et al. A combined use of microscopic traffic simulation and extreme value methods for traffic safety evaluation. Transportation Research Part C: Emerging Technologies, vol. 90, pp. 281-291, May 2018.
\bibitem{c23} Zheng L, Sayed T, Tageldin A. Before-after safety analysis using extreme value theory: a case of left-turn bay extension. Accident Analysis \& Prevention, vol. 121, pp. 258-267, October 2018.
\bibitem{c24} Zheng L, Sayed T. A full Bayes approach for traffic conflict-based before?after safety evaluation using extreme value theory. Accident Analysis \& Prevention, vol. 131, pp. 308-315, July 2019.







\end{thebibliography}
\end{document}